\documentclass[
,showpacs ,amssymb,superscriptaddress,aps
]{revtex4}
\usepackage{graphicx}
\usepackage{color}
\input{epsf}

\usepackage{amsmath,amssymb}
\usepackage{bm}
\usepackage{times}

\newcommand{\dalm}{\kern1pt\vbox{\hrule height 0.9pt\hbox{\vrule width 0.9pt
\hskip 2.5pt\vbox{\vskip 5.5pt}\hskip 3pt\vrule width 0.3pt}\hrule height 0.3pt}
\kern1pt}

\begin{document}



\title{Probing mass-radius relation of protoneutron stars from gravitational-wave asteroseismology}

\author{Hajime Sotani}
\email{hajime.sotani@nao.ac.jp}
\affiliation{Division of Theoretical Astronomy, National Astronomical Observatory of Japan, 2-21-1 Osawa, Mitaka, Tokyo 181-8588, Japan}

\author{Takami Kuroda}
\affiliation{Department of Physics, University of Basel, Klingelbergstrasse 82, 4056 Basel, Switzerland}
\affiliation{Institut f\"ur Kernphysik, Technische Universit\"at Darmstadt, Schlossgartenstrasse 9, 64289 Darmstadt, Germany}

\author{Tomoya Takiwaki}
\affiliation{Division of Theoretical Astronomy, National Astronomical Observatory of Japan, 2-21-1 Osawa, Mitaka, Tokyo 181-8588, Japan}
\affiliation{Center for Computational Astrophysics, National Astronomical Observatory of Japan, 2-21-1 Osawa, Mitaka, Tokyo 181-8588, Japan}

\author{Kei Kotake}
\affiliation{Department of Applied Physics, Fukuoka University, 8-19-1, Jonan, Nanakuma, Fukuoka, 814-0180, Japan}
\affiliation{Max Planck Institut f\"ur Astrophysik, Karl-Schwarzschild-Str. 1, 85748, 
Garching, Germany}

\date{\today}

\begin{abstract}
The gravitational-wave (GW) asteroseismology is a powerful technique for 
extracting interior information of compact objects.
In this work, we focus on spacetime modes, the so-called $w$-modes, of 
GWs emitted from 
a proto-neutron star (PNS) in the postbounce phase of core-collapse supernovae.
 Using results from recent three-dimensional supernova models,
 we study how to infer the properties of the PNS based on a
 quasi-normal mode analysis in the context of the GW asteroseismology.
We find that the $w_1$-mode 
frequency multiplied by the PNS radius is expressed as a linear function 
with respect to the ratio of the PNS mass to the PNS radius. This relation is 
 insensitive to the nuclear equation of state (EOS) employed in this work.
 Combining with another universal relation of the $f$-mode oscillations,
 we point out that the time dependent mass-radius relation of the PNS can be 
obtained by observing both 
the $f$- and $w_1$-mode GWs simultaneously. Our results suggest that 
the simultaneous detection of the two modes could provide a new probe into finite-temperature nuclear EOS that predominantly 
 determines the PNS evolution.
\end{abstract}

%
\maketitle
\section{Introduction}
\label{sec:I}

At last, the first direct detection of gravitational waves (GWs) was made
 by the twin detectors of the Laser Interferometer Gravitational-Wave Observatory (LIGO) 
from two binary black hole (BH) mergers \cite{GW1,GW2}. 
In addition to LIGO, second-generation detectors like Advanced 
VIRGO \cite{advv} and KAGRA \cite{aso13} will be operational in 
the coming years. Furthermore,  third-generation detectors like 
 Einstein Telescope (ET) and Cosmic Explorer (CE) are being proposed \cite{punturo,CE}. 
At such high level of precision, these detectors are sensitive enough to a 
wide variety of compact objects. The primary targets are compact binary 
coalescence such as the merger of BHs and/or neutron stars (NSs)
(e.g., \cite{schutzreview}). Other intriguing sources (e.g., \cite{nils}) include core-collapse supernovae (CCSNe) \cite{KotakeGWreview}, which 
mark the catastrophic end of massive stars 
and produce all these compact objects.

Extensive numerical simulations have been done so far to study GW signatures from 
core-collapse supernovae (e.g., \cite{Murphy09,MJM2013,CDAF2013,Yakunin15,Ott13,KKT2016,Andresen16}). It is now almost certain that the $g$-mode oscillations excited 
in the vicinity of the protoneutron star (PNS) are one of the most strong GW emission 
processes in the postbounce supernova core. The typical GW frequency ($f_{g}$) of 
the $g$-mode is approximately expressed as $f_g \sim G M_{\rm PNS}/ R_{\rm PNS}^2$ 
\cite{Murphy09,MJM2013,CDAF2013,KKT2016} where $G$ is the gravitational constant,
 $M_{\rm PNS}$ and $R_{\rm PNS}$ represent the mass and radius of the PNS, respectively.
 Predominantly due to the mass accretion to the PNS, $f_g$ increases with time after bounce \cite{Murphy09,MJM2013}.
 Neutrino-driven convection and the standing accretion-shock instability (SASI)
\cite{Blondin03,Foglizzo06} play a key role to effect the activity of the mass 
accretion to the PNS. 
 For progenitors with high-compactness \cite{evanott}, the SASI 
 is more likely to dominate over neutrino-driven convection 
 in the accretion phase \cite{Hanke13,Nakamura15}. In such a case, 
large-scale anisotropic flow associated with the SASI leads to strong GW 
emission, whose typical 
GW frequency closely matches with that of the SASI motion \cite{KKT2016,Andresen16}.
 The SASI-induced GW frequency $f_{\rm SASI}\sim100$ Hz is significantly lower 
 than that of the $g$-mode frequency ($f_{g}\sim500$-1000 Hz). The detection of 
 these distinct GW features is thus expected to provide a smoking-gun evidence to 
 infer which one is more dominant in the supernova engine, neutrino-driven 
convection or the SASI \cite{KKT2016,Andresen16}.

The linear perturbation approach (e.g., \cite{KS1999} for a review)
 is another way, which enables us to study 
 the fundamental properties of compact objects sometimes in a simplified manner.
 With the quasi-normal mode analysis,
 one can determine 
 the oscillation frequencies, once a background model is prepared. 
Since the oscillation spectra strongly depend on the properties of the source,
 one may extract the information of the source object via the correlation between the oscillation spectra and stellar properties. This technique is known as asteroseismology. 
In fact, important properties of the NS physics such as 
nuclear symmetry energy in the crust have been constrained by observations of 
 quasi-periodic oscillations in the magnetar giant flares \cite{SW2009,GNHL2011,SNIO2012,SNIO2013a,SNIO2013b,SIO2016,SIO2017}. It has been also suggested that the properties 
of a cold NS, such as the mass ($M$), radius ($R$), and the nuclear equation of
 state (EOS), 
could be constrained by the direct observations of GWs (e.g., \cite{AK1996,AK1998,STM2001,SH2003,SYMT2011,PA2012,DGKK2013}). 

 Among the above studies, it has been shown that the frequencies of the fundamental 
oscillations ($f$-modes) and of the spacetime oscillations ($w$-modes) from 
 cold NSs are characterized by the square root of the stellar average 
density, $(M/R^3)^{1/2}$, and the stellar compactness, $M/R$, respectively,
 independently from the EOS \cite{AK1996,AK1998}. 
Thus, if simultaneous observations of the $f$- and $w_1$-modes in GWs are made 
possible, one could in principle determine the mass and the radius of the cold NS
 from the average density and the compactness (see \cite{KS1999} for a review). 
We remark that, in order to determine the EOS for a high density region, one 
 may need to detect the GWs from several cold NSs to sample the mass-radius relation.

Unlike a cold NS, the perturbative analyses in the case of a PNS 
are only a few \cite{FMP2003,FKAO2015,ST2016,Camelio17}. 
This may come from the difficulty for providing the background model for the PNS. That is, the structure of the PNS depends on not only the relation between the pressure and energy density but also the radial profile of the electron fraction ($Y_e$) and entropy per baryon ($s$), while the distribution of $Y_e$ and $s$ are determined only 
via neutrino radiation-hydrodynamics core-collapse supernova simulations that are 
generally computationally expensive \cite{jankarev,bernhardrev,kotakerev}. 
In our previous study \cite{ST2016}, we focused on the $f$-modes from the PNS, 
adopting the $Y_e$ and $s$ distributions from one-dimensional (1D) 
supernova simulations. Then, we have shown that the $f$-mode GW frequency
  is characterized by the average density of the PNS independently of the progenitor 
models. 

In this work, we focus on another specific oscillation from a PNS, i.e., $w$-modes. Unlike fluid oscillations, these spacetime modes can be considered only in the relativistic framework \cite{wmode,wmode1}. $w$-modes are oscillations of spacetime itself, which are almost independent of the fluid oscillations. It is also known that the oscillation spectra of the axial-type $w$-modes are quite similar to those of the polar-type $w$-modes \cite{AKK1996}. Thus, in this paper, we will consider the axial-type $w$-modes from the PNS. Regarding the background model, we use results from 
  three-dimensional (3D) general-relativistic (GR) simulations in Ref. \cite{KKT2016}. 
By combining with our previous finding about the $f$-mode \cite{ST2016},
we investigate how we can enhance the predicative power of extracting 
the information of the PNS via the $w$-mode GWs using the outcomes of 
 most recent 3D supernova models. 

 This paper is structured as follows. In Sec. \ref{PNSmodel}, we describe
 the PNS models that we use as a background in this work.
 We then briefly summarize the 
 perturbation equations for a quasi-normal mode analysis in Sec. \ref{sec:III}. 
The main results are presented in Sec. \ref{sec:IV}. We give a conclusion in Sec. \ref{sec:V}.
Unless otherwise mentioned, 
we adopt geometric units in the following,
$c=G=1$, where $c$ denotes the speed of light, and the metric signature is $(-,+,+,+)$.

\section{PNS Models}
 \label{PNSmodel}
 Regarding our background models of the PNS, we take results 
 from Ref. \cite{KKT2016}, where 3D-GR simulations \cite{kuroda12} have been done to follow the hydrodynamics from the onset of core collapse of a $15M_\odot$ star 
\cite{WW95}, through core bounce, up to $\sim$ 250 ms after bounce. 
We consider that the 3D model is more appropriate and realistic to describe
the PNS evolution particularly just after bounce.
This is simply because that hydrodynamics above the PNS surface
is far from spherically symmetric and it effects both SN explosion mechanism and thus PNS thermodynamics.
Two EOSs were used with different nuclear interaction treatments, 
which are SFHx \cite{SHF2013} and TM1 \cite{HS2010}. 
In the following, the two 3D-GR models are named SFHx and TM1,
which simply reflects the EOS employed.
For SFHx and TM1, the maximum gravitational mass ($M_{\rm max}$) and 
the radius ($\bar{R}$) in the vertical part of the mass-radius relationship of a cold NS are
 $M_{\rm max}=2.13$ and $2.21M_\odot$, and $\bar{R}= 12$ and $14.5$ km, 
respectively \citep{Fischer14}. Thus SFHx is softer than TM1. 
Note that SFHx is not only the best-fit model with the observational mass-radius
relation of cold NSs \cite{SLB2010}, but also agrees much better than TM1 
with respect to nuclear and neutron matter constraints on the EOS \cite{Tsang2012}.
Both EOSs are compatible with the $2M_{\odot}$ NS mass measurements 
 \citep{Demorest10,antoniadis}. 
 
Hydrodynamic evolutions are rather common between SFHx and TM1, which is 
characterized by the prompt convection phase shortly after bounce
($T_{\rm pb}\alt 20$ ms with $T_{\rm pb}$ representing the postbounce time),
 then the linear (or quiescent) phase ($20 \alt T_{\rm pb}\alt 100$ ms),
 which is followed by the non-linear phase where the strong SASI dominates over 
  neutrino-driven convection in the postshock region
($100 \alt T_{\rm pb}\alt 300$ ms).
The softer EOS (SFHx) makes 
 the PNS radii and the shock at the shock-stall more compact
 compared to TM1. This leads to more stronger activity of the (sloshing and
 spiral) SASI motion in SFHx
 compared to TM1 (see \cite{KKT2016} for more 
details). 

\begin{figure*}[hbt]
\begin{center}
\begin{tabular}{cc}
\includegraphics[scale=0.5]{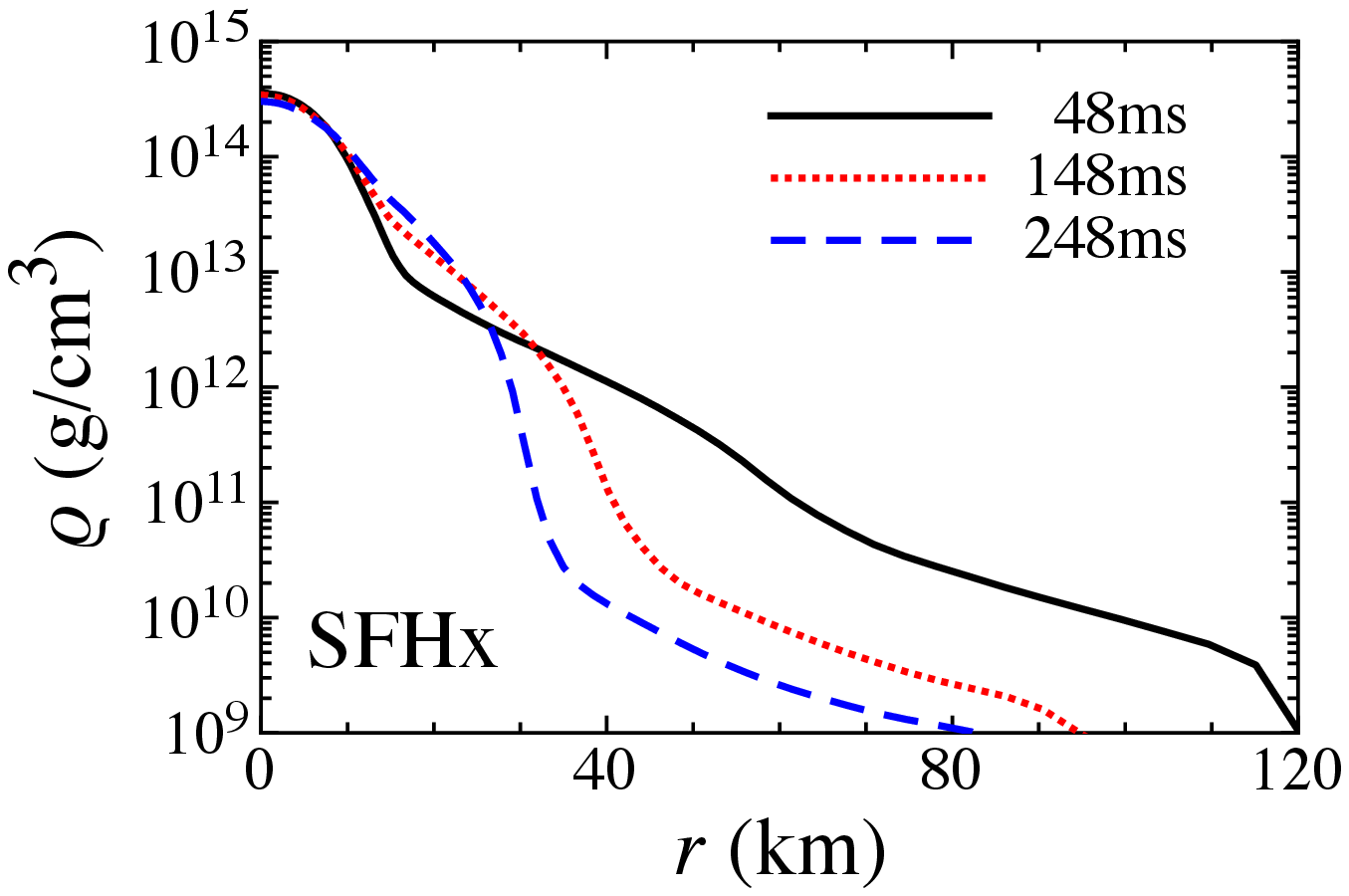} &
\includegraphics[scale=0.5]{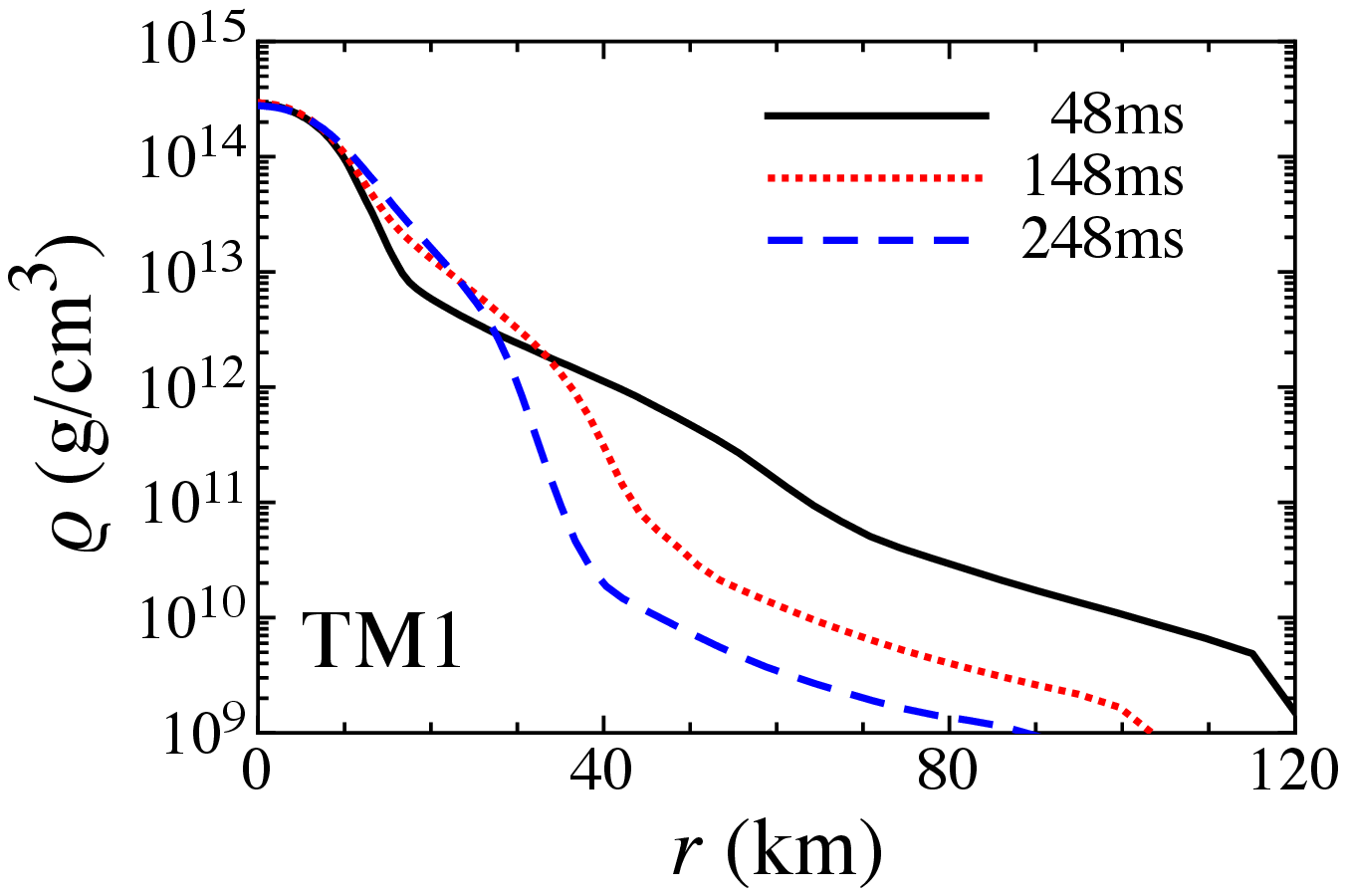}
\end{tabular}
\end{center}
\caption{
(Spherically-averaged) radial profiles of the rest-mass density at 48, 148, and 248 ms after core bounce. The left and right panel corresponds to SFHx and TM1, respectively.
}
\label{fig:rho-r}
\end{figure*}

  Fig. \ref{fig:rho-r} shows radial profiles of the rest-mass density at 
three representative epochs after bounce ($T_{\rm pb} = 48$, 148, and 248 ms) for 
SFHx (left panel) and TM1 (right panel), respectively. Each timeslice 
corresponds to the linear phase 
($T_{\rm pb} = 48$ ms), the early ($T_{\rm pb} = 148$ ms) and late 
($T_{\rm pb} = 248$ ms) non-linear phase covered in the simulation, respectively 
(see also Fig.2 in \cite{KKT2016}). The maximum density for SFHx
(left panel, $\rho\agt2\times10^{14}$ g cm$^{-3}$) is a few 10\% higher 
  compared to TM1 (right panel). This is because SFHx is softer than TM1 as 
mentioned above. 
In fact, Fig. 2 shows that the PNS radius (left panel) is 
more compact for SFHx. Here the surface of the PNS is defined at a fiducial 
rest-mass density of $\rho_s=10^{10}$ g cm$^{-3}$, which is relatively 
lower in the literature (e.g., \cite{takiwaki14}), but necessary in order to 
include the nascent PNS from the 3D-GR models with limited simulation time
 after bounce.
In right panel, we plot gravitational mass of the PNS $M_{\rm PNS}$
(evaluated by Eq. (\ref{eq:Madm}) in Appendix \ref{sec:appendix_1}) for given spherically averaged hydro and metric datas.
We shortly mention the accuracy of $M_{\rm PNS}$ which is used later in our analysis.
Although the baryon mass conservation is strictly satisfied because of our conservative formula,
the gravitational mass is not conserved with the same accuracy in general
(the energy loss by gravitational waves is negligible for CCSNe) in the BSSN formalism.
The violation can be $\sim1\%$ in our code \citep{kuroda12}.
It is also not straightforward to estimate the gravitational mass of the PNS
with taking into account the non-negligible energy loss by neutrinos.
Furthermore we first take spherically average with a simple zeroth order spacial interpolation from 3D cartesian to 1D spherical coordinates
and afterward we evaluate $M_{\rm PNS}$.
Therefore the gravitational mass of the PNS can differ from its true value of the order of $\sim1 \%(\sim0.01$M$_\odot)$.
In Appendix \ref{sec:appendix_1}, we discuss impact of numerical accuracy in $M_{\rm PNS}$ for our results.

\begin{figure}[hbt]
\begin{center}
\begin{tabular}{cc}
\includegraphics[scale=0.5]{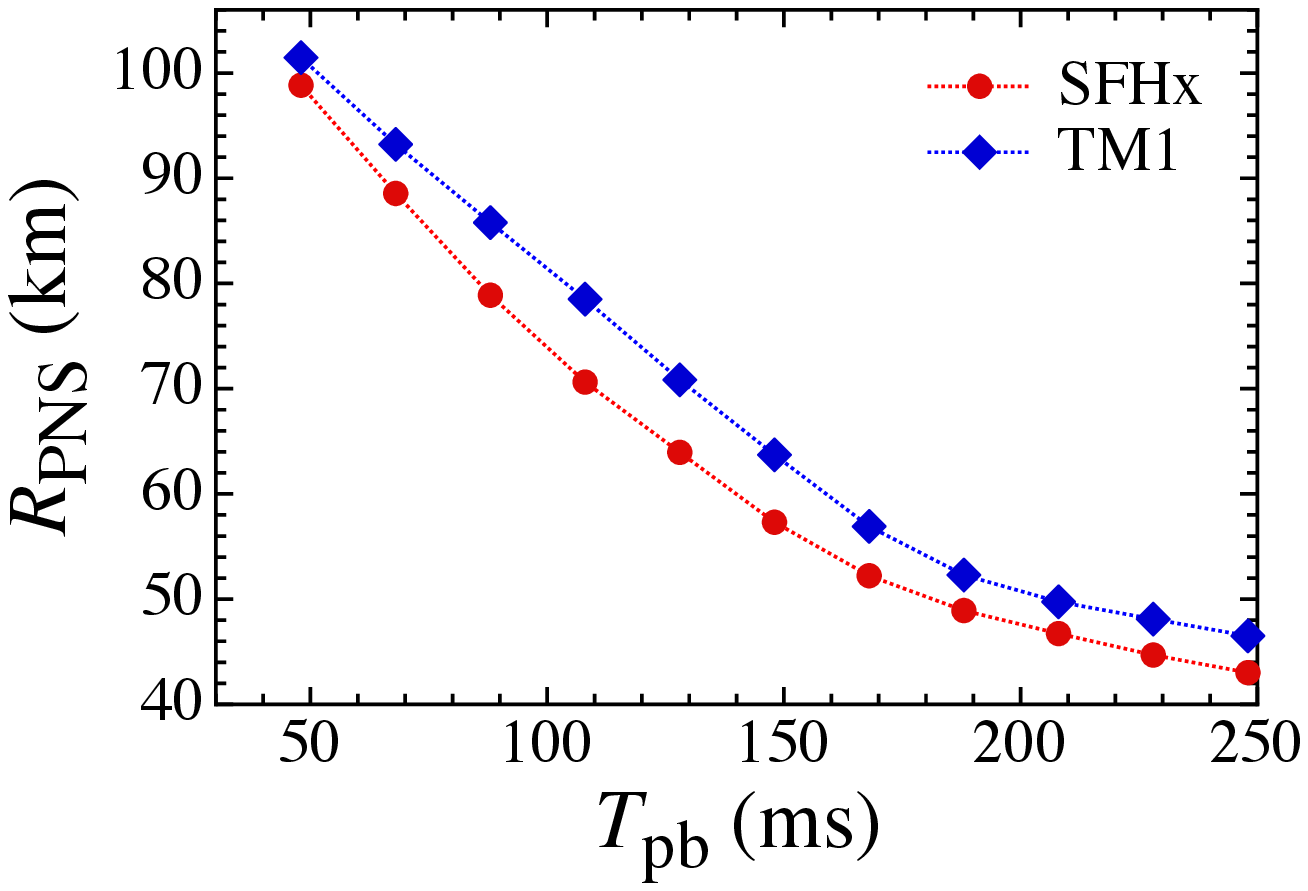} &
\includegraphics[scale=0.5]{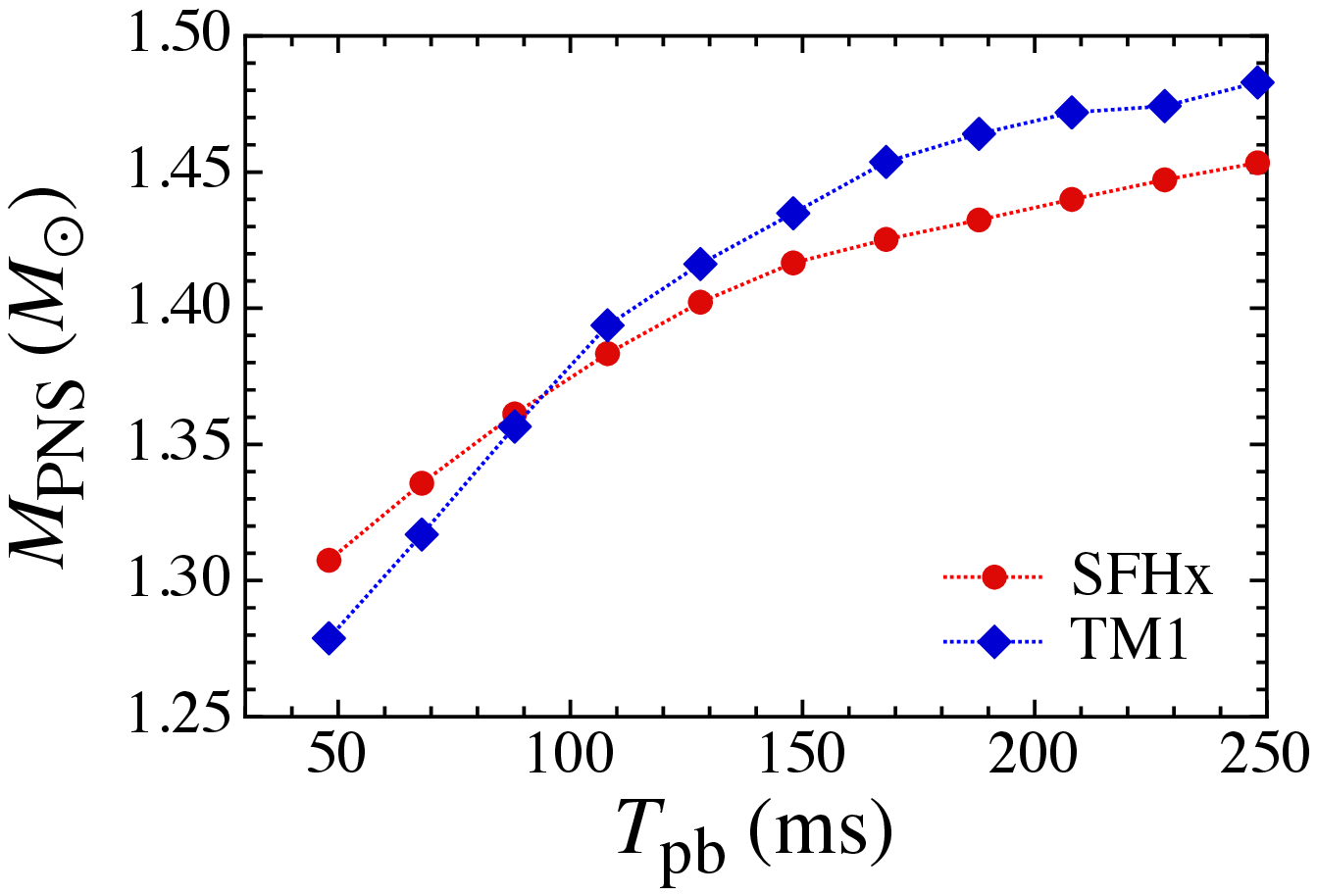}
\end{tabular}
\end{center}
\caption{
Time evolution of the PNS radius (left panel) and its gravitational mass (right panel) 
as a function of the postbounce time. The circles and diamonds corresponds to SFHx 
and TM1, respectively. The surface of the PNS is defined at a fiducial rest-mass density of $\rho_s=10^{10}$ g cm$^{-3}$. 
}
\label{fig:Mt}
\end{figure}

\begin{figure}[hbt]
\begin{center}
\begin{tabular}{cc}
\includegraphics[scale=0.5]{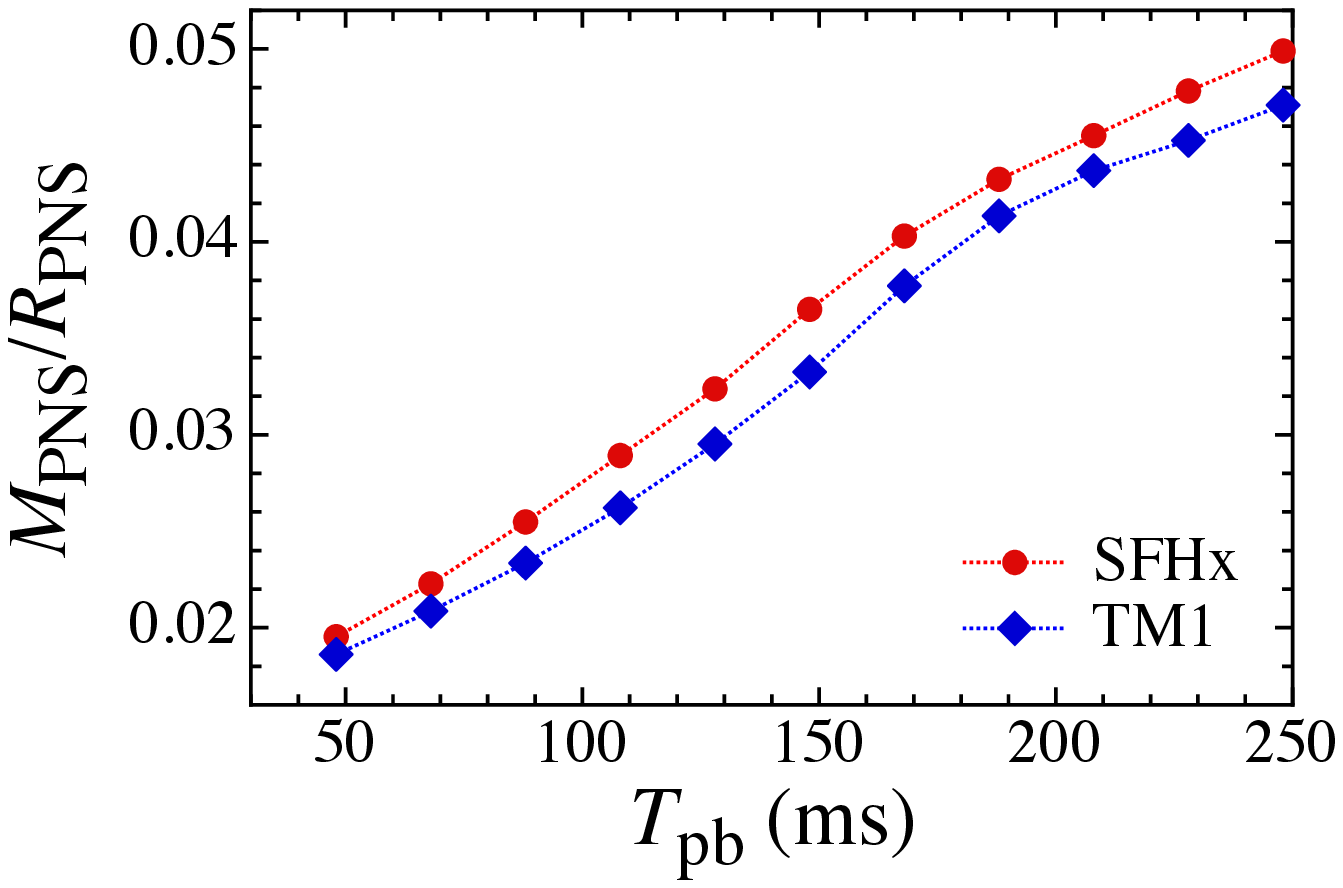} &
\includegraphics[scale=0.5]{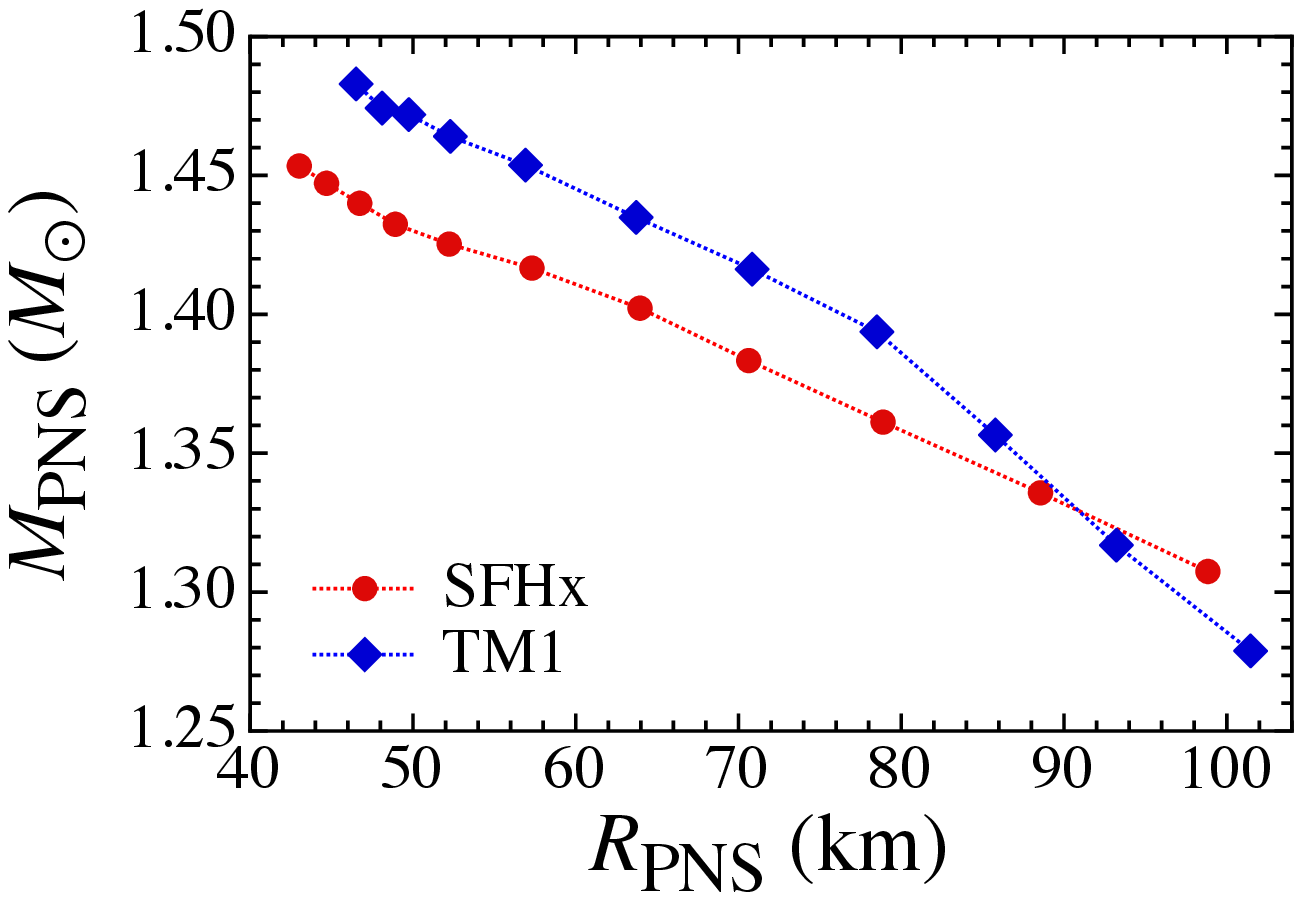} 
\end{tabular}
\end{center}
\caption{
Left: Same as Fig. \ref{fig:Mt}, but for the time evolution of the stellar compactness 
after bounce. Right: Sequences of the masses and radii of PNSs for SFHx and TM1. 
Note that the points at the left (smaller PNS radius) correspond to late 
postbounce phase, whereas the points at the right correspond to early phase (larger PNS radius).
}
\label{fig:MR_MRt}
\end{figure}

The left panel of Fig. 3 shows the evolution of 
the ``compactness" of the PNS that is defined 
 by $M_{\rm PNS}/R_{\rm PNS}$ for SFHx (red line) and TM1 (blue line).
 As one would imagine, the compactness of the PNS is higher for SFHx compared to TM1
even after we consider the inaccuracy of $\sim1\%$ in $M_{\rm PNS}$.
The right panel of Fig. 3 depicts the time evolution of $M_{\rm PNS}$ as a function 
of $R_{\rm PNS}$. 
The PNS with the softer EOS (SFHx) evolves 
 from larger to smaller PNS radius with bigger to smaller enclosed mass compared 
 to the stiffer EOS (TM1). 
Depending on the stiffness of the EOSs, one can see that the evolution track in the $M_{\rm PNS}-R_{\rm PNS}$ plane differs significantly.

To extract the metric from the background models in a suitable form, we perform 
the following coordinate transformation. 
 In the background models obtained by numerical relativity simulation
 (e.g., \cite{KKT2016}), the line element is given as
\begin{equation}
ds^2=-\alpha^2dt^2+\gamma_{ij}(dx^i+\beta^idt)(dx^j+\beta^jdt),
\label{eq:ds_BSSN}
\end{equation}
where $\alpha$, $\beta^i$, and $\gamma_{ij}$ are the lapse, shift vector, and three metric, 
respectively.
 If one assumes that the hydrodynamical background is static and spherically 
symmetric, the spacetime in the isotropic coordinates can also be written as
\begin{equation}
  ds^2 =-\frac{\left(1-\frac{M}{2\hat{r}}\right )^2}{\left(1+\frac{M}{2\hat{r}}\right )^2} dt^2 + \left(1+\frac{M}{2\hat{r}}\right )^4 
  (d\hat{r}^2+\hat{r}^2d\theta^2+\hat{r}^2\rm{sin}^2\theta d\phi^2),
\label{eq:ds_isotropic}
\end{equation}
where $\hat{r}$ and $M$ denote the isotropic radius $\hat{r}=\sqrt{x^2+y^2+z^2}$ 
and the enclosed gravitational mass, respectively.
From Eqs. (\ref{eq:ds_BSSN}) and (\ref{eq:ds_isotropic}), 
one can easily check the validity of our static and spherically symmetric 
background assumption by comparing $\gamma_{\hat{r}\hat{r}}$ and 
$(1+M/2\hat{r})^4$ (see Appendix \ref{sec:appendix_1} for detail).

Next, we perform coordinate transformation from the isotropic, i.e. Eqs. (\ref{eq:ds_BSSN}) or (\ref{eq:ds_isotropic}), to the following spherically symmetric 
spacetime,
\begin{equation}
  ds^2 =-e^{2\Phi} dt^2 + e^{2\Lambda} dr^2 + r^2\left(d\theta^2 + \sin^2\theta d\phi^2\right),
\label{eq:ds_spherical}
\end{equation}
where $\Phi$ and $\Lambda$ are functions of only $r$.
This metric is similar to the Schwarzschild metric and we apply the well-known conversion relation $r=\hat r(1+M/2\hat r)^2$.
In addition, $\Lambda$ is associated with the mass function $M$ in such a way that $e^{-2\Lambda}=1-2M/r$. With this metric form, the four-velocity of fluid element is given by $u^\mu=(e^{-\Phi},0,0,0)$.

\section{Perturbation equations for axial $w$-mode gravitational waves}
\label{sec:III}

On the PNS models mentioned in the previous section, we examine the oscillations and its spectra with the linear perturbation approach. In particular, when one focuses on axial type oscillations, the metric perturbation, $h_{\mu\nu}$, with the Regge-Wheeler gauge can be decomposed as
\begin{equation}
 h_{\mu\nu} =
 \sum_{\ell=2}^{\infty} \sum_{m=-\ell}^{\ell}\left(
 \begin{array}{cccc}
 0  &  0 & - h_{0,\ell m} {\sin^{-1}\theta} \partial_{\phi} & h_{0,\ell m} \sin\theta \, \partial_{\theta} \\
 \ast & 0 & -h_{1,\ell m} {\sin^{-1}\theta} \partial_{\phi} & h_{1,\ell m} \sin\theta \, \partial_{\theta} \\
 \ast & \ast & 0 & 0 \\
 \ast & \ast & 0 & 0 \\
 \end{array}
 \right) Y_{\ell m}\,,
 \end{equation}
where $Y_{\ell m}$ is the spherical harmonics with the angular indexes 
$\ell$ and $m$, noting that $h_{0,\ell m}$ and $h_{1,\ell m}$ are functions of $t$ and $r$ \cite{KS1999}. Additionally, the perturbation of the four-velocity is given by 
\begin{equation}
  \delta u^\mu = \sum_{\ell=2}^{\infty} \sum_{m=-\ell}^{\ell}\left(0,0,-\frac{\delta u_{\ell m}}{r^2\sin\theta}\partial_{\phi}Y_{\ell m}, \frac{\delta u_{\ell m}}{r^2\sin\theta}\partial_{\theta}Y_{\ell m}\right),
\end{equation}
while the perturbations of pressure and energy density should be zero for axial type oscillations.

The perturbation equation governing the axial type of GWs on the spherically symmetric background can be expressed as a single wave equation \cite{TC1967,CF1991}, such as 
\begin{equation}
  -\frac{\partial^2 X_{\ell m}}{\partial t^2} + \frac{\partial^2 X_{\ell m}}{\partial r_*^2} 
     - e^{2\Phi}\left[\frac{\ell(\ell+1)}{r^2} - \frac{6m}{r^3} + 4\pi(\varepsilon - p)\right]X_{\ell m}=0, \label{eq:wave}
\end{equation}
where $X_{\ell m}$ is related to the metric perturbation, $h_{1,\ell m}$, via $rX_{\ell m}= e^{\Phi - \Lambda} h_{1,\ell m}$, while $r_*$ is the tortoise coordinate defined as $r_*=r+2M\ln (r/2M-1)$. That is, $\partial_r = e^{\Lambda - \Phi}\partial_{r_*}$. The remaining variables, $h_{0,\ell m}$ and $\delta u_{\ell m}$, can be computed with $h_{1, \ell m}$ from the relations $\partial_t h_{0,\ell m} = e^{\Phi-\Lambda}X_{\ell m} + r\partial_{r_*} X_{\ell m}$ and $\delta u_{\ell m}= -e^{-\Phi}h_{0, \ell m}$. We remark that Eq. (\ref{eq:wave}) outside the star reduces to the well-known Regge-Wheeler equation. Hereafter, we omit the index of $(\ell, m)$ for simplicity.

In fact, by solving this system one can obtain the specific oscillation spectra of 
GWs, i.e., the so-called $w$-modes \cite{wmode,wmode1}. Replacing $X_{\ell m}$ in Eq. (\ref{eq:wave}) with $X_{\ell m}(t,r) = X(r)\exp(i\omega t)$, one gets the perturbation equation with respect to the eigenvalue $\omega$,
\begin{equation}
  X'' + \left(\Phi' - \Lambda'\right)X' + e^{2\Lambda}\left[\omega^2 e^{-2\Phi} - \frac{\ell(\ell+1)}{r^2} 
       + \frac{6m}{r^3} - 4\pi(\varepsilon - p)\right]X=0.
\end{equation}
By imposing appropriate boundary conditions, the problem to solve becomes the eigenvalue problem. The boundary conditions are the regularity condition at the stellar center and the outgoing wave condition at spatial infinity.

The eigenvalue $\omega$ becomes a complex number, because GWs carry out the oscillation energy, where the real and imaginary parts of $\omega$ correspond to the oscillation frequency ($f={\rm Re}(\omega)/2\pi$) and damping rate ($1/\tau={\rm Im}(\omega)$), respectively, where $\tau$ corresponds to the damping time of each mode. To determine such a complex frequency, we adopt the continuous fractional method proposed by Leaver \cite{Leaver}.

\section{Asteroseismology with $w$-modes}
\label{sec:IV}

The spacetime modes ($w$-modes) have two families, i.e., $w_{\rm II}$ and ``ordinary" $w$-modes \cite{wmode,wmode1}. As shown in Appendix \ref{sec:appendix_2}, for cold  NSs, a few $w_{\rm II}$-modes are excited, whose damping rate (${\rm Im}(\omega)$) is larger than its oscillation frequency (${\rm Re}(\omega)$). On the other hand, infinite number of $w$-modes can exist, which are referred to as $w_1$, $w_2$, $\cdots$, $w_n$-modes in order from the lowest oscillation frequency. So, in the similar way to cold NSs, we identify the spacetime modes with ${\rm Re}(\omega)$ larger than ${\rm Im}(\omega)$ as the ``ordinary" $w$-modes for PNSs. Hereafter, the ``ordinary" $w$-modes are called just as the $w$-modes.

\begin{figure*}
\begin{center}
\begin{tabular}{cc}
\includegraphics[scale=0.5]{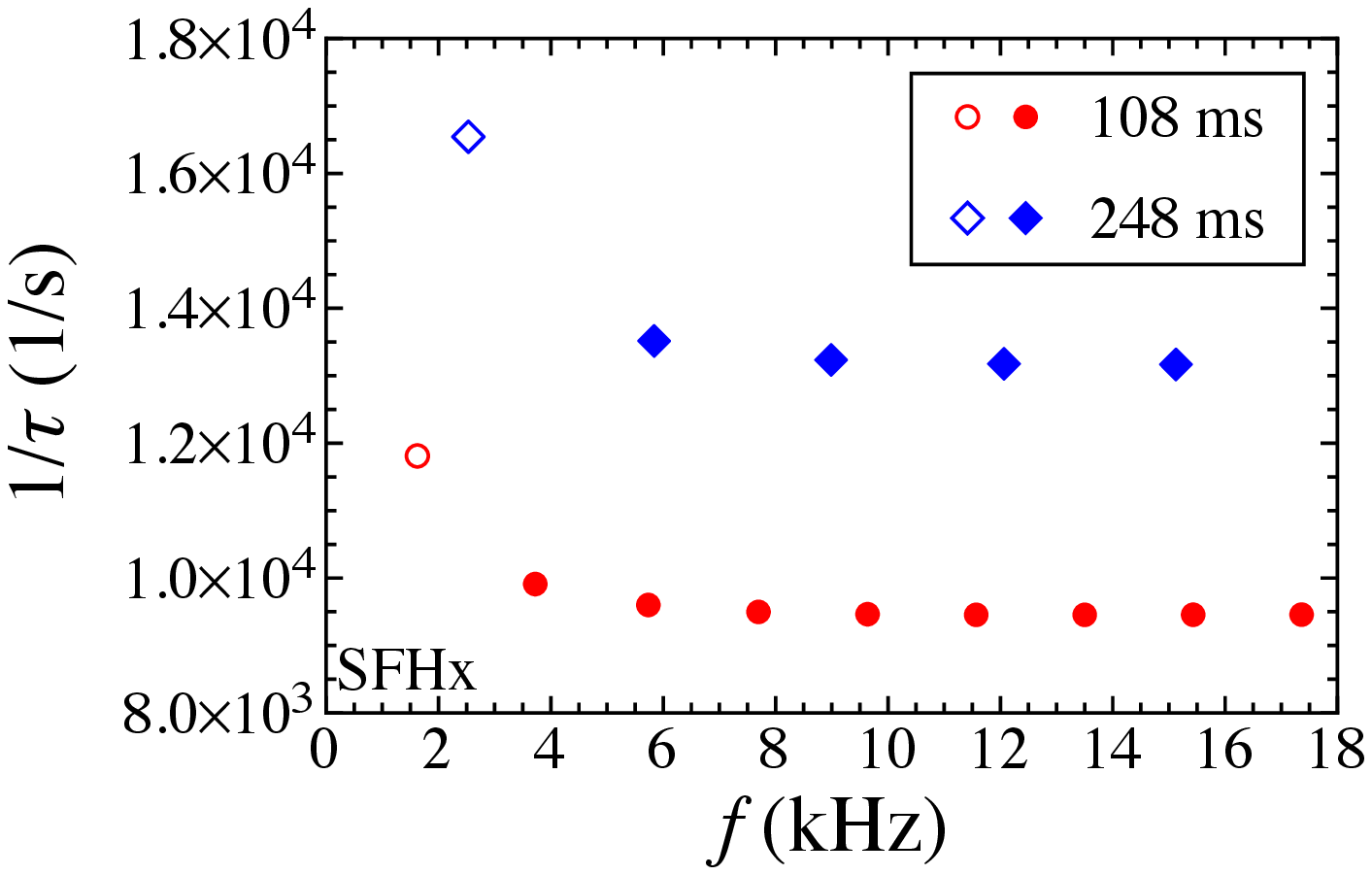} &
\includegraphics[scale=0.5]{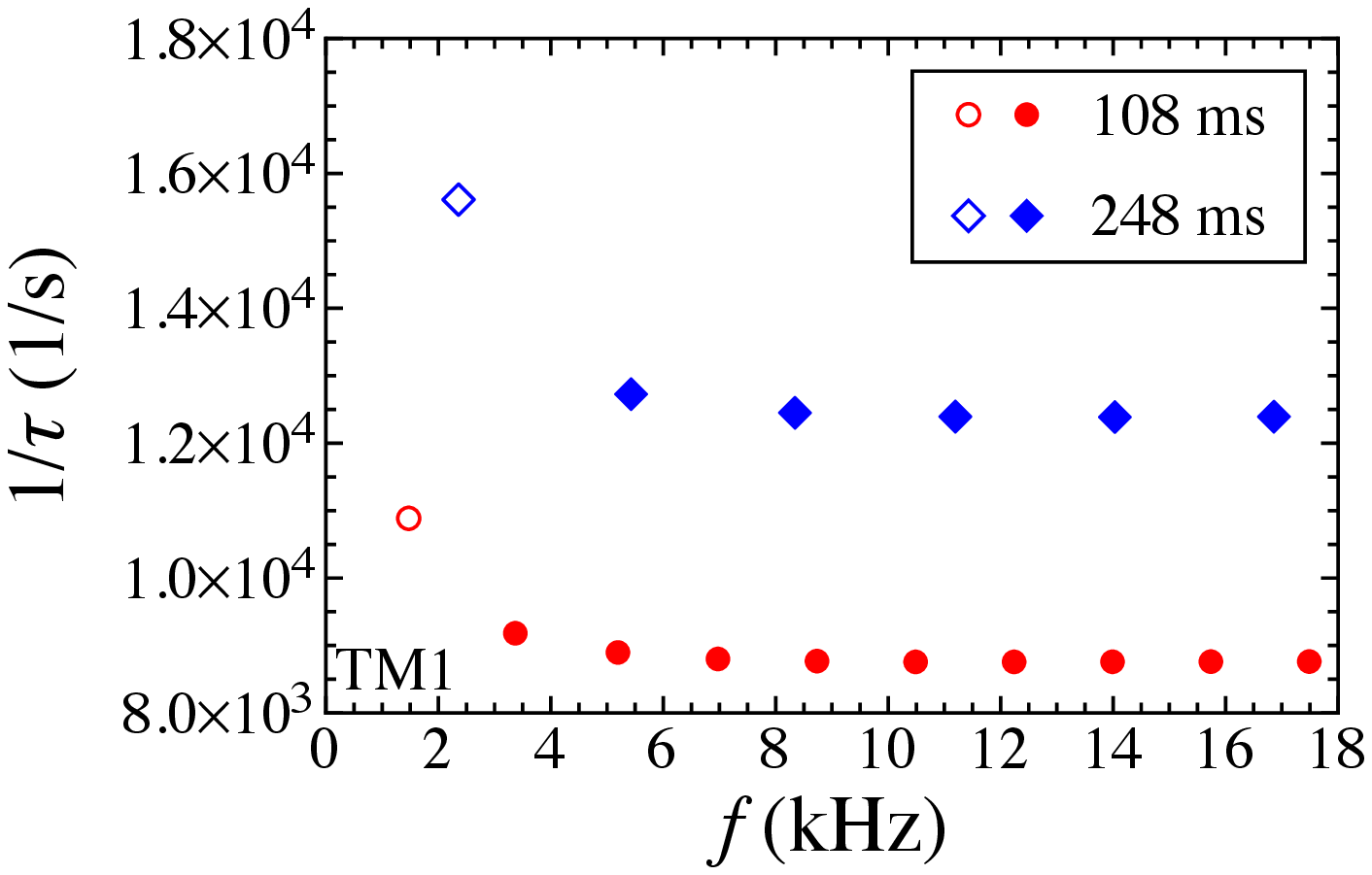}
\end{tabular}
\end{center}
\caption{
Frequency and damping rate of the axial spacetime modes for PNSs. The left and right panels correspond to the results for SFHx and TM1 EOSs, respectively, where the circles and diamonds are shown for the PNS models at 108 and 248 ms after core bounce. The open and solid marks correspond to the $w_{\rm II}$ and ``ordinary" $w$-modes.
}
\label{fig:wmode}
\end{figure*}

\begin{figure*}
\begin{center}
\begin{tabular}{cc}
\includegraphics[scale=0.5]{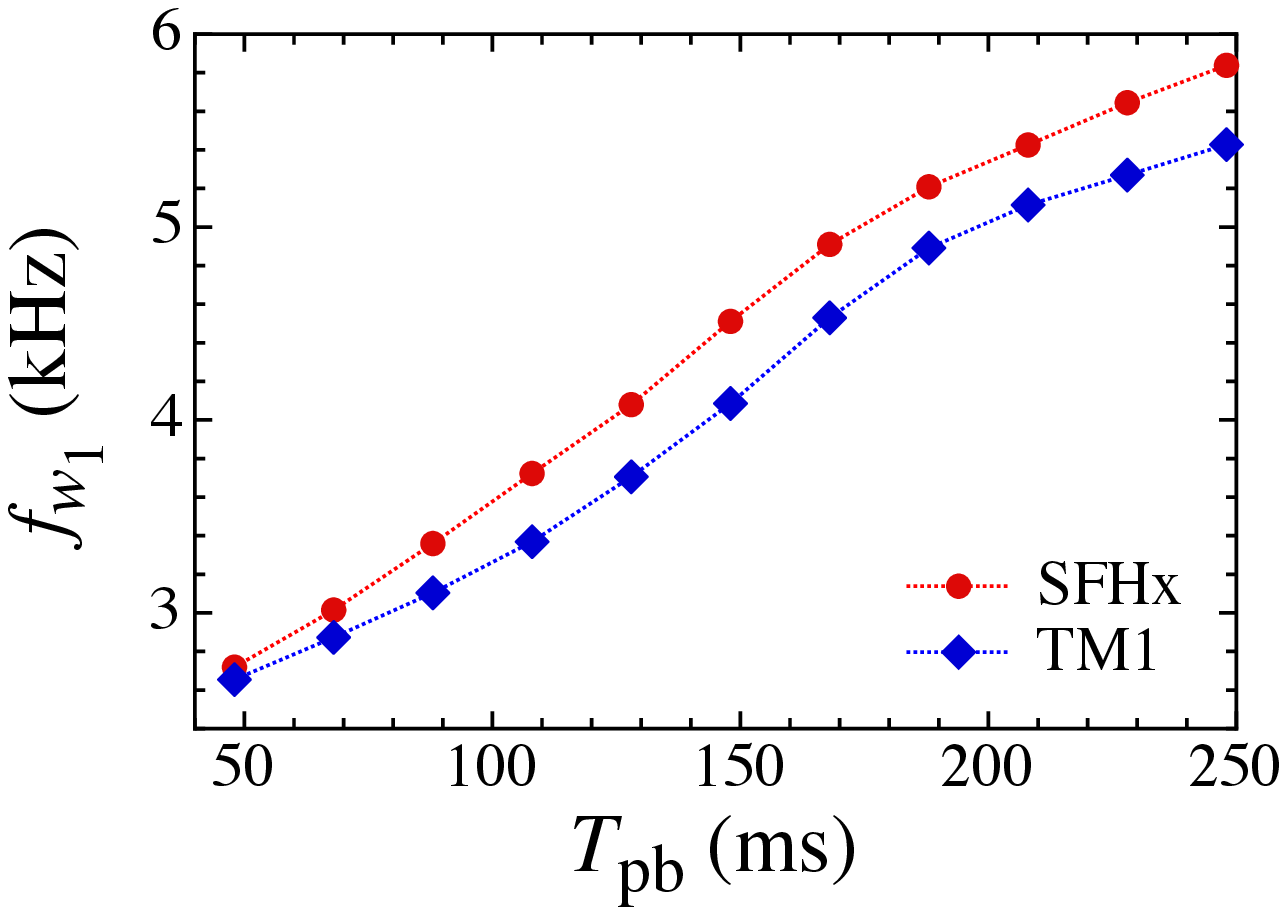} &
\includegraphics[scale=0.5]{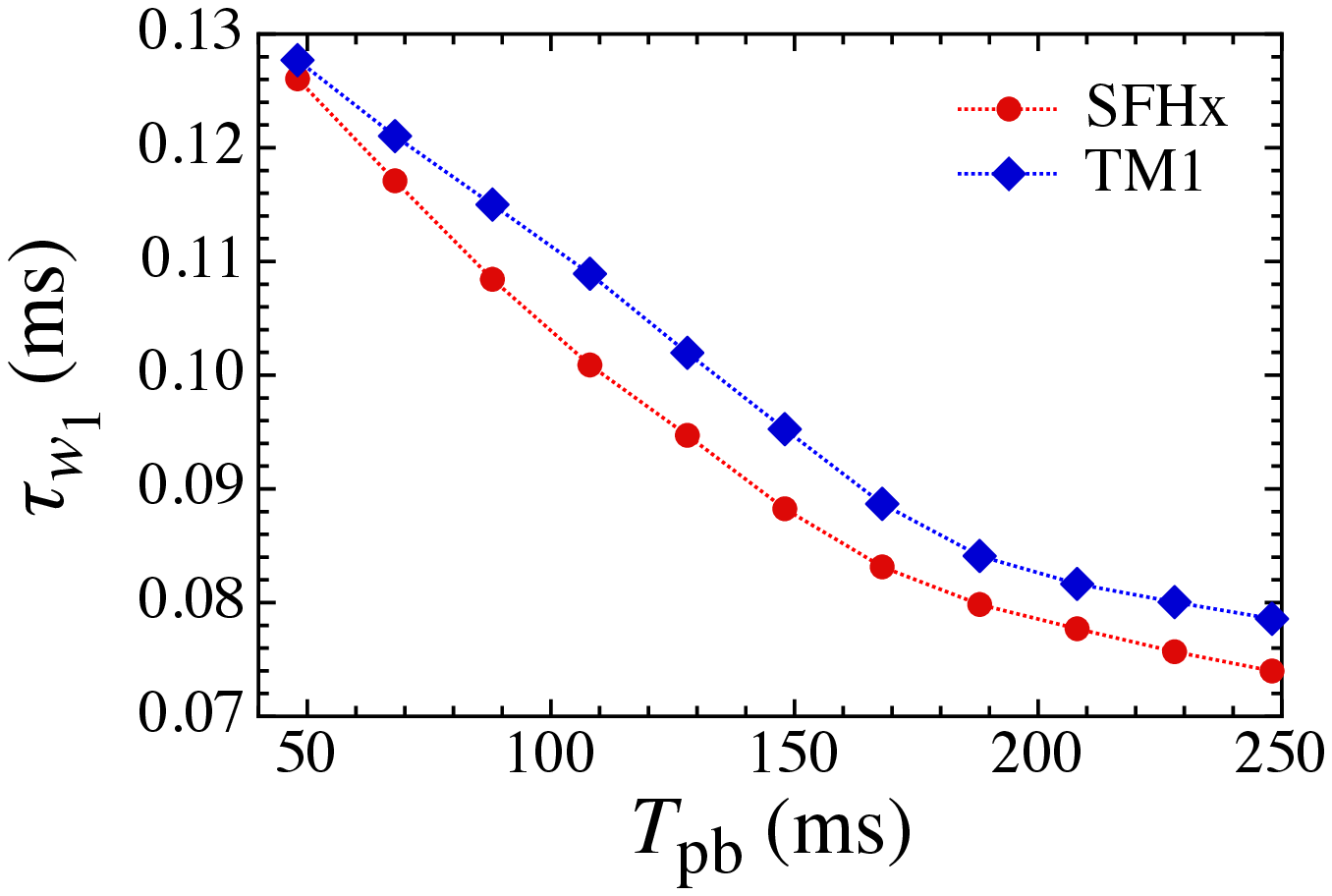}
\end{tabular}
\end{center}
\caption{
Evolutions of frequency $(f_{w_1})$ and damping time $(\tau_{w_1})$ for the $w_1$-mode. The circles and diamonds correspond to SFHx and TM1, respectively.
}
\label{fig:wmode-t}
\end{figure*}

In Fig. \ref{fig:wmode}, we show the frequency and damping rate of the axial 
spacetime modes for the PNS models at the two postbounce times of $T_{\rm pb} =$ 
108 ms (circles) and 248 ms (diamonds), where the left and right panels correspond
 to the results with SFHx and TM1 (EOS). 
In this figure, the open marks denote the $w_{\rm II}$-modes, while the solid marks denote the $w$-modes. Thus, the leftmost solid marks correspond to the $w_1$-mode (fundamental $w$-mode) for each PNS model. From this figure, one can observe that the damping rate of $w_n$-mode is almost constant independently of the index $n$, which is different behavior from the case of cold NSs as shown in Fig. \ref{fig:ColdNSw}. In fact, the damping rate of $w_n$-modes increase with the index $n$ for cold NSs. 
With respect to the $w_1$-mode (Fig. \ref{fig:wmode-t}), we show the time evolution of the frequency ($f_{w_1}$) and damping time ($\tau_{w_1}$) as a function of 
postbounce time for SFHx and TM1, respectively. We remark that the damping time is the time with which the GW amplitude reduces by $1/e$. In the early phase of $w_1$-mode oscillations of PNSs, the frequency is only a few kHz, which is significantly smaller than that for cold NSs, while the damping time is around 0.1 ms, which is much larger than that for cold NSs. This is good news from the observational point of view. The direct detection of such a frequencies with the future (or even current) 
GW detectors might be possible, depending on the radiation energy of $w_1$-mode and the distance to a source object.

It is known that the frequency of $w_1$-mode for cold NSs can be characterized by the stellar compactness. That is, Andersson and Kokkotas have shown that for cold NSs
 the $w_1$-mode frequencies multiplied by the stellar radius are characterized by the stellar compactness independently of the EOS of neutron star matter \cite{AK1998}, such as
\begin{equation}
  f_{w_1}^{\rm (NS)}\ {\rm (kHz)} \approx \left[20.92-9.14\left(\frac{M}{1.4M_\odot}\right)\left(\frac{R}{10\ {\rm km}}\right)^{-1}\right]
      \left(\frac{R}{10\ {\rm km}}\right)^{-1}.  \label{eq:w1_CNS}
\end{equation}
This behavior comes from that the $w$-modes are oscillations of spacetime itself, which is almost independent from the matter oscillations. In the same way, the additional universal relation between the frequency of $f$-mode and stellar average density for cold NSs have been also derived  \cite{AK1998}, such as
\begin{equation}
  f_{f}^{\rm (NS)}\ {\rm (kHz)} \approx 0.78 + 1.635\left(\frac{M}{1.4M_\odot}\right)^{1/2}\left(\frac{R}{10\ {\rm km}}\right)^{-3/2}.
\end{equation}
This means that, via the simultaneous observations of the frequencies of $f$- and $w_1$-modes, one can get two different information about the compact object, which enables us to constrain the mass and radius of source object. This is an original idea  proposed by \cite{AK1998} to adopt the GW asteroseismology to the cold NSs. In this work, we revisit this in the context of the PNS, i.e., we will consider the possibility for obtaining the mass and radius of PNSs via the observations of the $f$- and $w_1$-modes GWs.

\begin{figure}
\begin{center}
\includegraphics[scale=0.5]{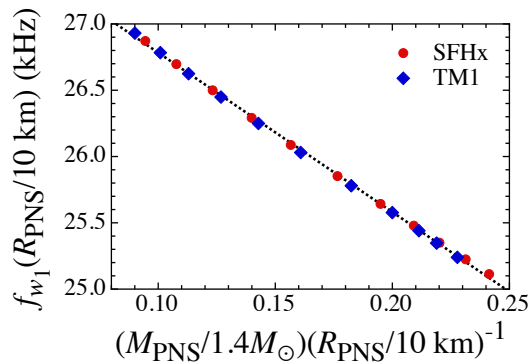}
\end{center}
\caption{
The $w_1$-mode frequencies multiplied by the normalized radius $f_{w_1}(R/10\ {\rm km})$ are shown as a function of normalized compactness $(M_{\rm PNS}/1.4M_\odot)(R_{\rm PNS}/10\ {\rm km})^{-1}$, where the circles and diamonds denote the results for SFHx and TM1 EOSs, respectively. The dotted line is a fitting formula given by Eq. (\ref{eq:w1_PNS}).
}
\label{fig:fR-MR10}
\end{figure}

We find that the similar universal relation for $w_1$-mode can be held even for the PNSs. In Fig. \ref{fig:fR-MR10}, we show the $w_1$-mode frequencies multiplied by the radius as a function of the compactness, where the circles and diamonds correspond to the results for SFHx and TM1, respectively. As shown in Fig. \ref{fig:MR_MRt}, since the compactness increases with time, the left side in Fig. \ref{fig:fR-MR10} corresponds to the early phase of PNSs. From this figure, we derive the fitting formula such as
\begin{equation}
  f_{w_1}^{\rm (PNS)}\ {\rm (kHz)} \approx \left[27.99-12.02\left(\frac{M_{\rm PNS}}{1.4M_\odot}\right)\left(\frac{R_{\rm PNS}}{10\ {\rm km}}\right)^{-1}\right] \left(\frac{R_{\rm PNS}}{10\ {\rm km}}\right)^{-1}. \label{eq:w1_PNS}
\end{equation}
We remark that the $w_1$-mode frequencies for PNSs expected from this fitting formula are significantly different from those for cold NSs expected from Eq. (\ref{eq:w1_CNS}), because the radius and mass of PNSs are different from those for cold NSs.
We also remark that the scaling law for PNSs with using the mass and radius is the same as that for cold NSs, but the coefficients in the law are different. So, the coefficients in the scaling law would vary with time and eventually approach the values for cold NSs. This would suggest that long-term GW astroseismology and the GW detection could potentially bridge the gap of the two formulae evolving from a PNS phase into a cold NS phase.

\begin{figure*}
\begin{center}
\begin{tabular}{ccc}
\includegraphics[scale=0.42]{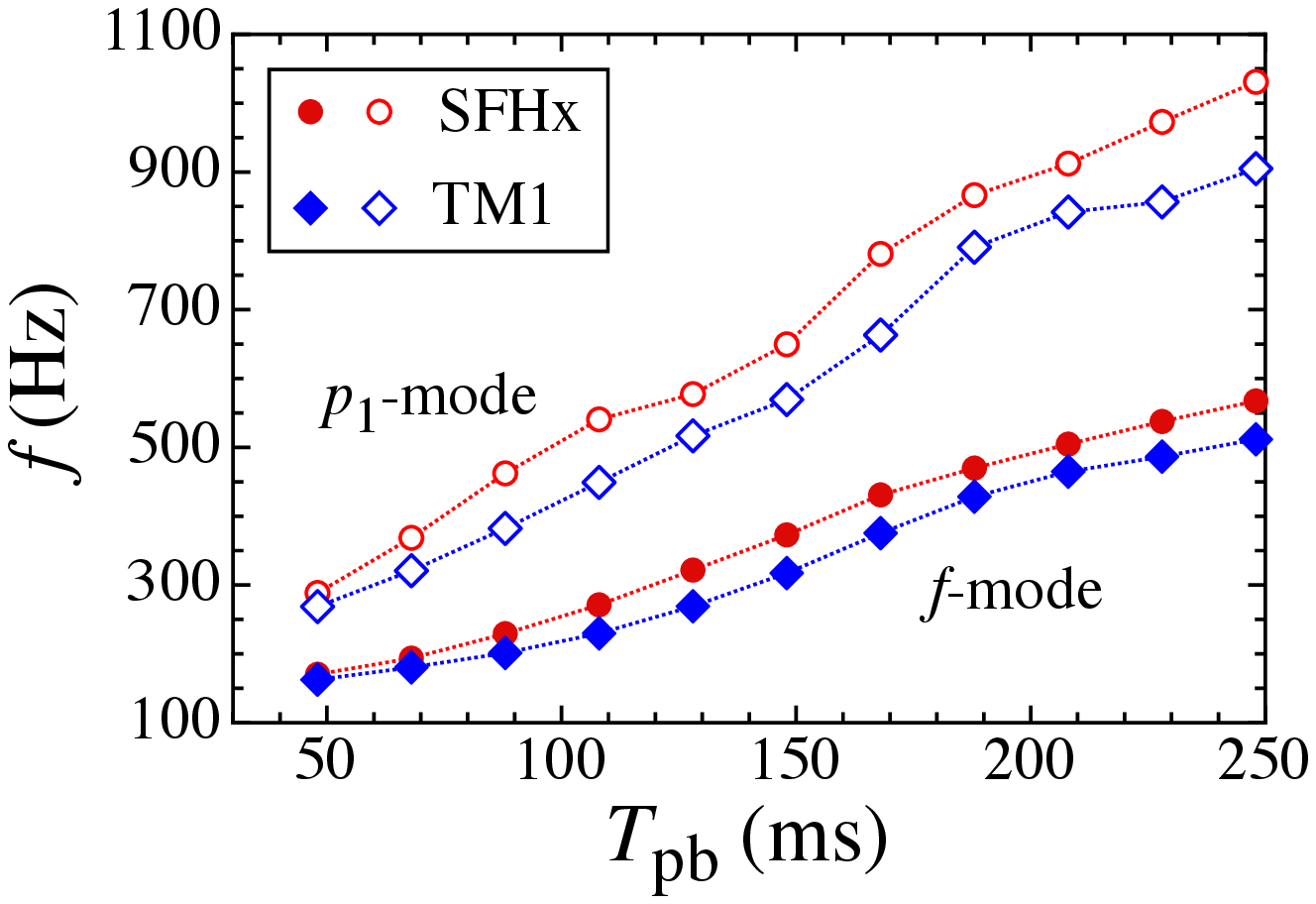} &
\includegraphics[scale=0.42]{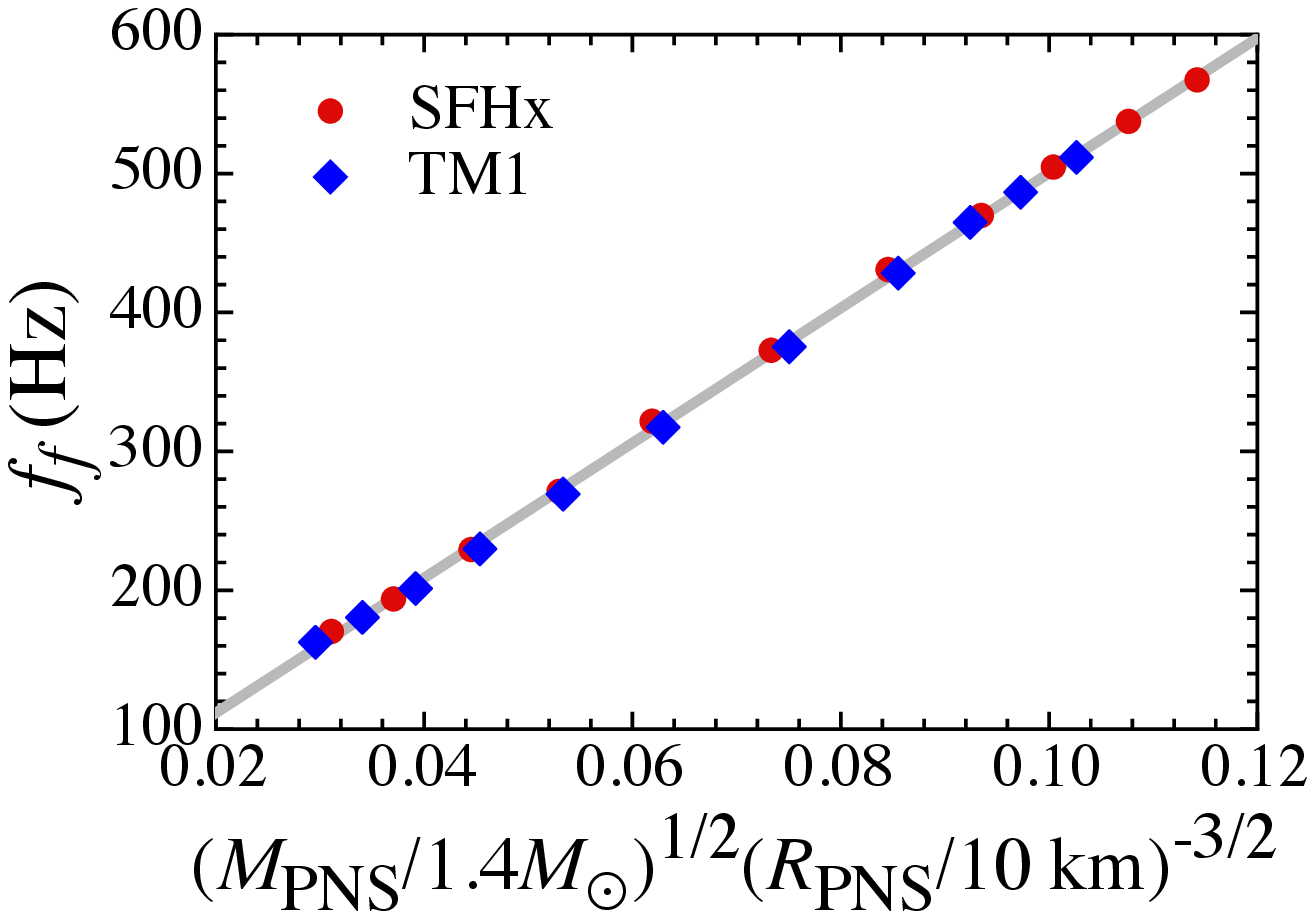} &
\includegraphics[scale=0.42]{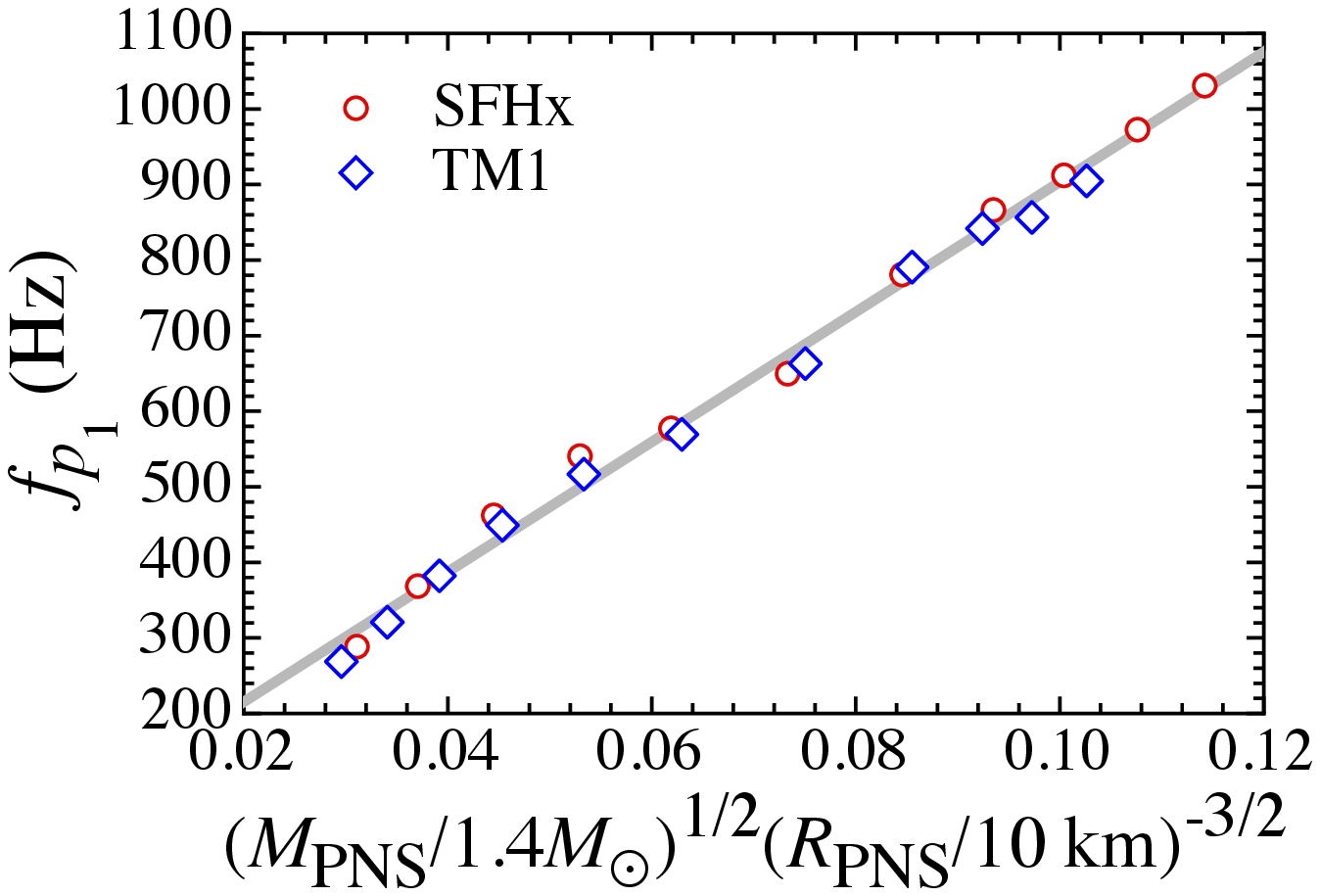} 
\end{tabular}
\end{center}
\caption{
Evolutions of $f$- and $p_1$-modes in GWs from PNSs after core bounce are shown in the left panes. The solid and open marks correspond to the $f$- and $p_1$-modes, while the circles and diamonds are respectively the results for SFHx and TM1. The middle and right panels shows respectively the frequencies of the $f$- and $p_1$-modes as a function of average density of PNSs. The solid line denotes the linear fitting given by Eqs. (\ref{eq:f_PNS}) and (\ref{eq:p1_PNS}).
}
\label{fig:fmode}
\end{figure*}

With respect to the $f$-mode on PNSs, we have derived the universal relation between the $f$-mode frequency and the average density of PNS independently of the progenitor models \cite{ST2016}. However, in order to consistently discuss the $f$-mode oscillations with the results of $w_1$-mode, we re-calculate by using the PNS models adopted in this paper with the same procedure as in \cite{ST2016}, i.e., with Cowling approximation neglecting the variation of entropy. Then, we get the time evolutions of $f$- and $p_1$-modes for SFHx and TM1 as shown in the left panel of Fig. \ref{fig:fmode}. It should be noticed that even $p_1$-mode frequency might be possible to observe because the frequencies in the early phase of PNS are only a few hundred Hz. In the same way as shown in \cite{ST2016}, we also confirm that the frequencies of $f$-modes can be expressed as a linear function of the average density of PNS independently of the adopted EOS (see the right panel of Fig. \ref{fig:fmode}), such as
\begin{align}
  f_{f}^{\rm (PNS)}\ {\rm (Hz)} &\approx 14.48 + 4859\left(\frac{M_{\rm PNS}}{1.4M_\odot}\right)^{1/2}
      \left(\frac{R_{\rm PNS}}{10\ {\rm km}}\right)^{-3/2},  \label{eq:f_PNS}  \\
  f_{p_1}^{\rm (PNS)}\ {\rm (Hz)} &\approx 43.29 + 8602\left(\frac{M_{\rm PNS}}{1.4M_\odot}\right)^{1/2}
      \left(\frac{R_{\rm PNS}}{10\ {\rm km}}\right)^{-3/2},  \label{eq:p1_PNS}
\end{align}
where the coefficients in the linear fits are modified a little from the previous one because the surface density of the PNS models adopted in this paper is different from that in \cite{ST2016}. In practice, these linear fits are also shown in the middle and right panels of Fig. \ref{fig:fmode} with solid lines. We remark that the frequencies of the $f$- and $p_1$-modes are the same dependence on the properties of PNSs, i.e., one can get only the information about the average density of PNS even if one will simultaneously detect the $f$- and $p_1$-modes.

Consequently, one can obtain the information of two different properties, which are combinations of $M_{\rm PNS}$ and $R_{\rm PNS}$, via Eqs. (\ref{eq:w1_PNS}) and (\ref{eq:f_PNS}) (or via Eqs. (\ref{eq:w1_PNS}) and (\ref{eq:p1_PNS})), if one would simultaneously detect the $f$- and $w_1$-modes (or the $p_1$- and $w_1$-modes) in GWs from PNSs, which enables us to know the values of $M_{\rm PNS}$ and $R_{\rm PNS}$. Furthermore, unlike the GW asteroseismology for cold NSs, for PNSs one might get the sequence in $M_{\rm PNS}-R_{\rm PNS}$ plain as shown in Fig. \ref{fig:MR_MRt} with the time evolution of the GW spectra from the PNS produced by just one supernova explosion, because $M_{\rm PNS}$ and $R_{\rm PNS}$ changes with time. Namely, in principle one would find the EOS via the detection of the GWs from just one supernova explosion.

Finally, we discuss the detectability of GWs from PNSs. In Refs. \cite{AK1996,AK1998}, the effective amplitude of $f$- and $w_1$-modes in GWs radiating from cold NSs are estimated, where the background stellar model should be static at least during the damping time. Since the damping time of $w_1$-mode from PNSs is typically $\tau_{w_1}\sim 0.1$ ms as shown in Fig. \ref{fig:wmode-t}, which is shorter than the typical timescale of change of PNS properties, one might possible to adopt the estimation of effective amplitude for $w_1$-mode derived in \cite{AK1996,AK1998} even for PNSs. On the other hand, if one estimates the damping time of $f$-mode for PNSs in the same way as for cold NSs, such as $\tau_f \sim R_{\rm PNS}^4/M_{\rm PNS}^3$ \cite{AK1998}, $\tau_f$ becomes $\sim 1-50$ second, which is much larger than the typical timescale of change of PNS properties. Thus, it must be inappropriate to adopt the estimation of effective amplitude for $f$-mode derived in \cite{AK1996,AK1998} in the case of PNSs. Thus, here we only consider the detectability of $w_1$-mode in gravitational waves. Even so, we may deduce that the upper limit of the effective amplitude of the $f$-mode in gravitational waves from PNSs would be around $h\sim 10^{-21}$, assuming that the $f$-mode oscillations can be also captured as well as the other excited modes in the previous numerical simulations of core-collapse supernovae \cite{CDAF2013,KKT2016,Andresen16}.

\begin{figure}
\begin{center}
\includegraphics[scale=0.5]{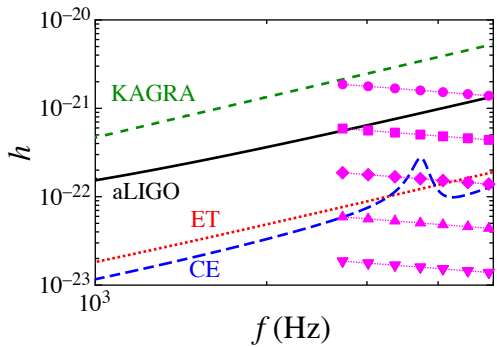}
\end{center}
\caption{
The effective amplitude of $w_1$-modes in gravitational waves radiated from the PNSs with SFHx EOS are shown together with the sensitivity curves of KAGRA, advanced LIGO (aLIGO), Einstein Telescope (ET), and Cosmic Explorer (CE). The circles, squares, diamonds, triangles, and upside-down triangles correspond to the results with $E_{\rm T}^{(w_1)}=10^{-4}M_\odot$, $10^{-5}M_\odot$, $10^{-6}M_\odot$, $10^{-7}M_\odot$, and $10^{-8}M_\odot$, respectively.
}
\label{fig:sensitivity}
\end{figure}

For PNSs, we choose that the energy of $w_1$-mode in the gravitational waves, $E_{w_1}$, for each time step, and estimate the effective amplitude of such gravitational waves with the same formula as in \cite{AK1996,AK1998}. Thus, the effective amplitude is given by 
\begin{equation}
  h_{\rm eff}^{(w_1)} \sim 7.7\times 10^{-23} \left(\frac{E_{w_1}}{10^{-10}\ M_\odot}\right)^{1/2}
       \left(\frac{4\ {\rm kHz}}{f_{w_1}}\right)^{1/2}\left(\frac{10\ {\rm kpc}}{D}\right),
\end{equation}
where $D$ denotes the distance between the source and the Earth. We remark that the effective amplitude depends on the frequencies of $w_1$-mode, which change with time. Assuming the total radiation energy with $w_1$-mode in the gravitational waves from PNS ($E_{\rm T}^{(w_1)}$), the energy for each time step ($E_{w_1}$) can be estimated as $E_{\rm T}^{(w_1)} \approx  E_{w_1} T_{w_1}/\tau_{w_1}$, where $T_{w_1}$ denotes the duration time of $w_1$-mode. In this paper, we simply assume that $T_{w_1}=250$ ms and $\tau_{w_1}=0.1$ ms. Since the total energy of $w_1$-mode in gravitational waves is also unknown, we consider five cases, i.e., $10^{-4}M_\odot$, $10^{-5}M_\odot$, $10^{-6}M_\odot$, $10^{-7}M_\odot$, and $10^{-8}M_\odot$, as the values of $E_{\rm T}^{(w_1)}$. Then, the expected effective amplitude of $w_1$-mode in gravitational waves radiated from PNSs with SFHx EOS is shown in Fig. \ref{fig:sensitivity} together with the sensitivity curves of KAGRA, advanced LIGO, Einstein Telescope, and Cosmic Explorer \cite{aso13,CE,CE2}. In this figure, the circles, squares, diamonds, triangles, and upside-down triangles denote for the cases with $E_{\rm T}^{(w_1)}=10^{-4}M_\odot$, $10^{-5}M_\odot$, $10^{-6}M_\odot$, $10^{-7}M_\odot$, and $10^{-8}M_\odot$, respectively. The leftmost marks of the effective amplitude for each mode correspond to the PNS model at 48 ms after core bounce, and the effective amplitude decreases with time. From this figure, the radiation energy of $E_{\rm T}^{(w_1)}=10^{-5}M_\odot$ seems to be marginal for the advanced LIGO.

\section{Conclusion}
\label{sec:V}

The GWs radiated from supernova explosions are one of the most promising 
sources. In this paper, we considered the GWs emitted from 
a PNS in the postbounce phase of core-collapse supernovae.
 In particular, we focused on the spacetime mode, 
the so-called $w$-mode. Regarding the background model, we used results from 
  most recent 3D-GR models. Then, we calculated the complex frequencies on 
such PNS models, assuming that the PNS model on each time step is static 
spherically symmetry. The real and imaginary parts of complex frequency 
correspond to the oscillation frequency and the damping rate.

We have found that the damping rate of $w_n$-modes for PNSs is almost
 independent from the index $n$, although that for cold NSs increases
 with $n$. Moreover, in the similar way to the case for cold NSs, we found
 that the $w_1$-mode frequency multiplied by the PNS radius can be expressed 
as a linear function of the compactness of PNSs independently of EOSs. 
The $w_1$-mode frequency of PNSs just after the core-bounce is typically 
around a few kHz, which might be better from the observational point of view. 
Using such a universal relation for $w_1$-mode frequency together with 
another universal relation for $f$-mode, where the frequency can be expressed 
as a linear function of the square root of the average density of PNSs independently of the 
progenitor models, one can get two different properties constructed with 
the mass and radius of the PNS, if one would detect simultaneously 
the both modes. Therefore, one would determine the mass and radius of PNSs 
in principle on each time step, which would enable us to study the 
finite-temperature EOS that predominantly determines the PNS evolution.

\acknowledgments
We are grateful to K. Hayama for providing the sensitivity curves for several 
GW detectors. This work was supported in part by Grant-in-Aid for Young 
Scientists (B) (Nos. 26800133, 17K14306) provided by JSPS, Grant-in-Aid for 
Scientific Research (C) (No. 17K05458) and (A) (No. 17H01130) provided by JSPS, and by Grants-in-Aid for Scientific Research on Innovative Areas through Grant (Nos. 15H00843, 15KK0173, 17H05206, 17H06357, 17H06364) provided by MEXT.

\appendix
\section{Validity of the static and spherically symmetric assumption for our 3D models}   
\label{sec:appendix_1}
\begin{figure*}
\begin{center}
\begin{tabular}{cc}
\includegraphics[width=75mm,clip,angle=0.]{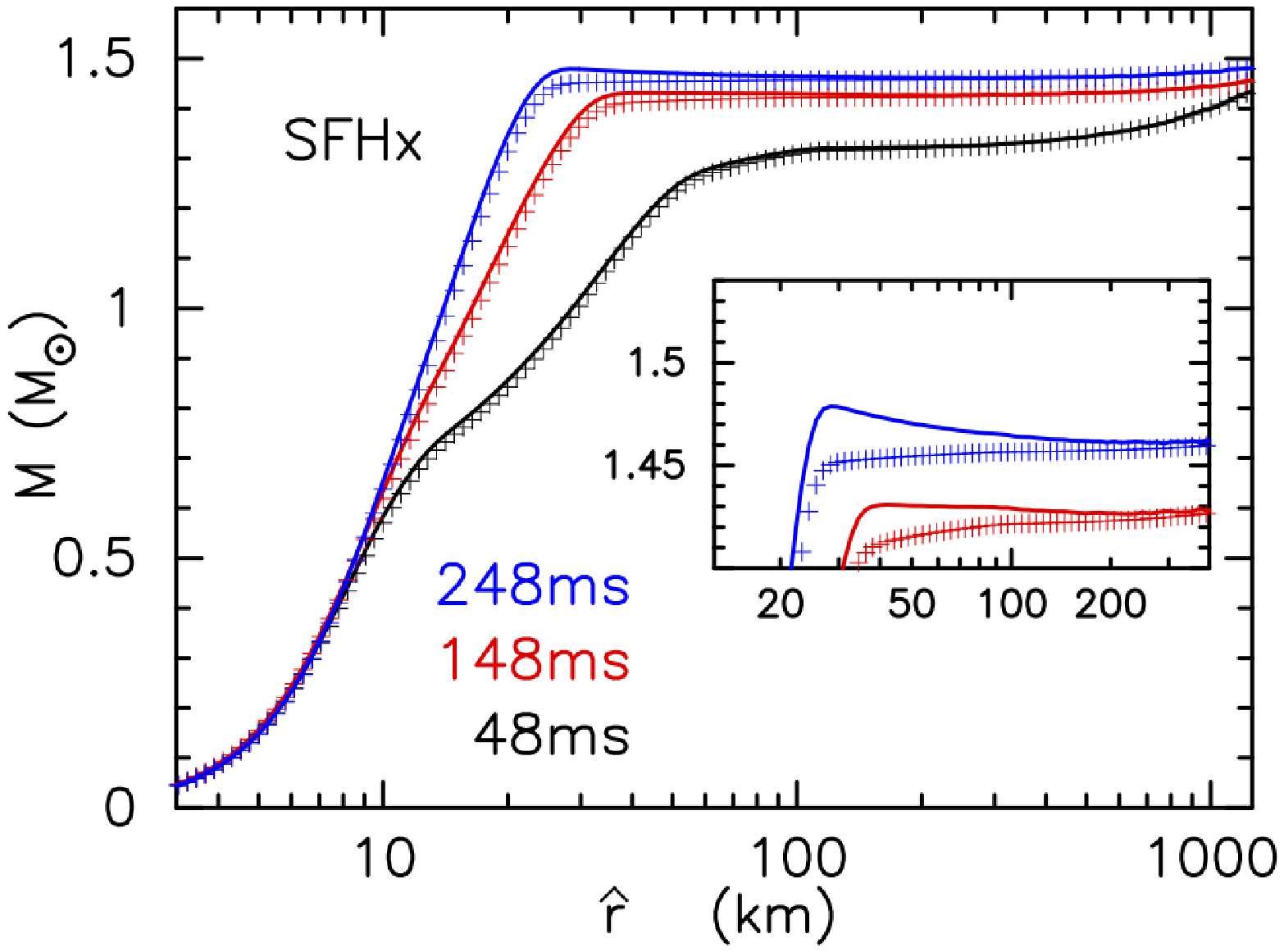}&
\includegraphics[width=75mm,clip,angle=0.]{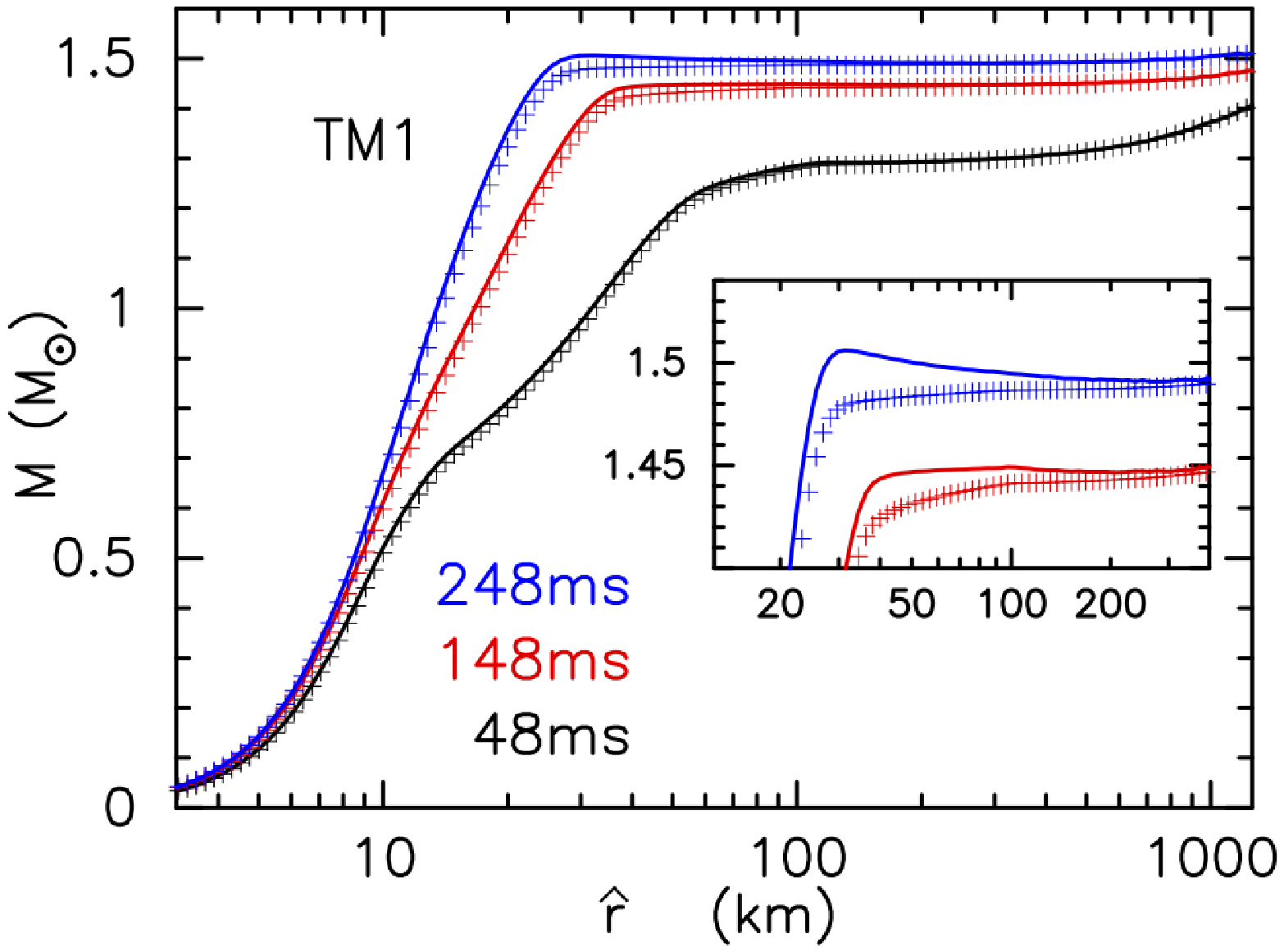}
\end{tabular}
\end{center}
\caption{
  The effective (solid) and actual (cross) gravitational mass as a function of the isotropic radius $\hat r$
 at 48, 148, and 248 msec after core bounce. The left and right panels correspond to the results with SFHx and TM1 EOSs, respectively.
}
\label{fig:Mgrv}
\end{figure*}

From Eqs. (\ref{eq:ds_BSSN}) and (\ref{eq:ds_isotropic}), one can check the validity of our static and spherically symmetric assumption by comparing $\gamma_{\hat{r}\hat{r}}$ and $(1+M/2\hat{r})^4$. In Fig. \ref{fig:Mgrv}, we plot the gravitational mass $M$ (cross) and the effective mass $2\hat{r}({\gamma_{\hat{r}\hat{r}}}^{1/4}-1)$ (solid line) at representative post bounce times $T_{\rm pb}=48$, 148, and 248 ms. The inner mini panel is a magnified view where the solid lines and crosses deviate the most. For $M$, we simply adopt the following ADM mass
\begin{eqnarray}
  \label{eq:Madm}
  M&=&\int dx^3
  \left[\rho_{\rm H} \psi^{5}
    +\frac{\psi^5}{16\pi}\Bigl(\tilde A_{ij}\tilde A^{ij}-\frac{2}{3}K^2-{\gamma}^{ij}\tilde{R}_{ij}\Bigr)
    \right],
\end{eqnarray}
where $\rho_{\rm H}=\rho hW^2-P$, with $\rho$, $h$, $W$, and $P$ being the rest mass density, enthalpy, Lorentz factor, and pressure, respectively, and $\psi$, $\tilde{A}_{ij}$, $K$, and $\tilde{R}_{ij}$ are the conformal factor, tracefree part of the extrinsic curvature, trace of the extrinsic curvature, and the Ricci tensor with respect to $\tilde{\gamma}_{ij}={\gamma}_{ij}\psi^{-4}$, respectively. Because of the highly convective motion within the shock radius $\hat{r}\alt100$ km, the effective gravitational mass calculated from $\gamma_{\hat{r}\hat{r}}$ does not match with $M$. The maximum deviation however is a few percent and we consider that it is sufficient to describe the background metric with $M$ instead of using simply $\gamma_{\hat{r}\hat{r}}$, since the effective gravitational mass at $\hat{r}\sim30-$50 km (solid line in the mini panel) has a negative gradient. 
In any way, we confirm that the resultant $w$-mode frequencies and damping rates with the actual mass we adopted in text are almost the same as those with the effective mass determined from $\gamma_{\hat{r}\hat{r}}$. For example, the difference in the $w_1$-mode frequency at 248 ms between the results with actual and effective masses is only $0.23 \%$ for SFHx and $0.14 \%$ for TM1, while the difference in the $w_1$-mode damping rate at 248 ms is only $0.98 \%$ for SFHx and $0.91 \%$ for TM1.

\section{$w$-mode for cold neutron stars}   
\label{sec:appendix_2}

\begin{figure}
\begin{center}
\includegraphics[scale=0.5]{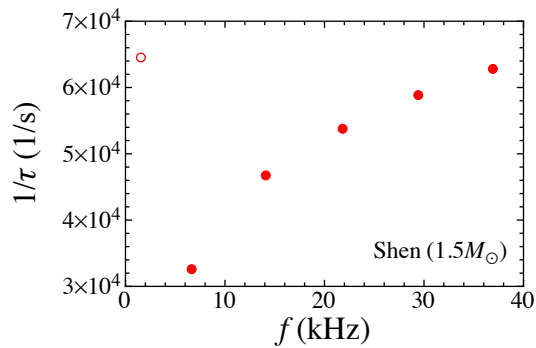}
\end{center}
\caption{
Typical frequency and damping rate of spacetime modes for a cold neutron star. This is a result for the neutron star with $1.5M_\odot$, which is constructed with the Shen EOS. The open and solid circles correspond to the $w_{\rm II}$ and ``ordinary" $w$-modes, respectively.
}
\label{fig:ColdNSw}
\end{figure}

For reference, in Fig. \ref{fig:ColdNSw} we show the complex frequencies of $w$-mode oscillations from a cold neutron star with $1.5M_\odot$ constructed with the Shen EOS \cite{Shen}. In this figure, the open circle corresponds to the $w_{\rm II}$-mode, while the solid circles are the $w$-modes. The $w$-modes are called as $w_1$, $w_2$, $\cdots$, $w_n$-modes from the lower frequencies. For the case of cold NSs, it is known that the damping rate of $w$-mode increases as the mode becomes higher-order oscillations \cite{wmode,wmode1}, as shown in this figure.

\end{document}